# $Fe_{0.79}Si_{0.07}B_{0.14}$ metallic glass gaskets for high-pressure research beyond 1 Mbar

Authors

**Weiwei Dong[a,b], Konstantin Glazyrin[a]\*, Saiana Khandarkhaeva[c], Timofey Fedotenko[c], Jozef Bednarčík[d], Leonid Dubrovinsky[e], Natalia Dubrovinskaia[c,f] and Hanns-Peter Liermann[a]**

[a]Deutsches Elektronen-Synchrotron DESY, Notkestr. 85, 22607 Hamburg, Germany

[b]Beijing Synchrotron Radiation Facility, Institute of High Energy Physics, Chinese Academy of Sciences, 19 Yuquan Rd, Beijing, 100049, People's Republic of China

[c]Material Physics and Technology at Extreme Conditions, Laboratory of Crystallography, University of Bayreuth, Universitätsstr. 30, Bayreuth, 95440, Germany

[d]Department of Condensed Matter Physics, Institute of Physics, P.J. Šafárik University, Šrobárova 1014/2, Košice, 041 54, Slovakia

[e]Bayerisches Geoinstitut, University of Bayreuth, Universitätsstr. 30, Bayreuth, 95440, Germany

[f]Department of Physics, Chemistry and Biology (IFM), Linköping University, Olaus Magnus väg, Linköping, SE-581 83, Sweden

Correspondence email: konstantin.glazyrin@desy.de

**Abstract**     A gasket is an important constituent of a diamond anvil cell (DAC) assembly, responsible for the sample chamber stability at extreme conditions for x-ray diffraction studies. In this work, we studied the performance of gaskets made of metallic glass $Fe_{0.79}Si_{0.07}B_{0.14}$ in a number of high-pressure x-ray diffraction (XRD) experiments in DACs equipped with conventional and toroidal-shape diamond anvils. The experiments were conducted in either uniaxial or radial geometry with x-ray beams of micron to sub-micron size. We report that the $Fe_{0.79}Si_{0.07}B_{0.14}$ metallic glass gaskets offered stable sample environment under compression exceeding one megabar in all XRD experiments described here, even in those involving inter- or small-molecule gases (e.g. Ne, $H_2$) used as pressure transmitting media or in those with laser heating in a DAC. These emphasize the material's importance for a great number of delicate experiments conducted under extreme conditions. Our results indicate that the application of $Fe_{0.79}Si_{0.07}B_{0.14}$ metallic glass gaskets in XRD experiments of both uniaxial and radial geometries substantially improves the signal-to-noise ratio in comparison to that with conventional gaskets made of Re, W, steel or other crystalline metals.

## 1. Introduction

New developments in the field of diamond anvil cell (DAC) technology have significantly increased the range of experimentally achievable pressures and temperatures providing new opportunities for high-pressure physics, chemistry, materials and planetary sciences. (Dubrovinsky et al., 2015, Loubeyre et al., 2020, Ji et al., 2019, Kono et al., 2016, Petitgirard et al., 2019). Modern large-scale facilities, such as synchrotron radiation (SR) and x-ray free electron laser (XFEL) enable the most sophisticated experiments at extreme conditions. The technical demands of the high-pressure research community, i.e. a sub-micron focal size with the highest possible flux, are gradually being addressed through the upgrade programs of the major synchrotron facilities to diffraction-limited storage rings (DLSRs), such as the ESRF-ECB (ESRF-EBS), APS-U (APS Upgrade), and PETRA IV (PETRA IV). In parallel, the high-pressure community contributes to the development, e.g. by improving signal-to-noise ratio (S/N) through the improvement of the DAC assembly.

Among numerous aspects related to the design of the DAC assembly, a choice of a gasket material takes the central part. The gasket containing a sample chamber should ideally consist of a material with high yield strength and good ductility to support contents of the sample chamber up to target pressures, which often exceed 1 Mbar. For this reason, gaskets are often made of strong metals such as Re, W or steel. However, these crystalline materials made from high-Z elements produce intense x-ray diffraction signal even if the gasket material is hit only by the strongly reduced intensity of focused beam 'tail'. Thus, most of conventional gasket materials may significantly contribute to the undesirable background scattering (on top of the Compton and diffuse scattering from the diamond anvils, etc.). These well-known contributions complicate the diffraction pattern analysis and lead to a degradation of the S/N ratio. For high-pressure experiments of uniaxial geometry, where the x-ray beam goes through the diamond anvils parallel to the DAC's compression axis (**Fig. 1a**), the 'parasitic' diffraction originating from the gasket material can be eliminated or considerably reduced, if the size of the x-ray spot is significantly smaller than the diameter of the sample chamber. In the case of the radial geometry, where the x-ray beam is perpendicular to the DAC's compression axis and x-rays go through the gasket (**Fig. 1b**), the diffraction from the gasket material is unavoidable. To reduce its contribution in the radial diffraction geometry, beryllium and epoxy mixtures containing low-Z elements or compounds are most commonly used. The properties of such materials are, however, limiting their application. For example, beryllium is toxic, and epoxies are often quite soft. Among hard materials, gaskets made of diamond (Zou *et al.*, 2001), c-BN (Funamori & Sato, 2008, Wang *et al.*, 2011), and amorphous boron epoxies (a-BE) (Lin *et al.*, 2003, Merkel & Yagi, 2005, Rosa *et al.*, 2016) have been efficiently used. Such light-element materials are very transparent for high energy x-rays, however, their sharp diffraction peaks produced by crystalline components of the epoxy mixtures (even diamond or c-BN has large scattering volume) can significantly overlap with the signal from the sample causing a deterioration of the data quality. Therefore, despite these great progresses in application of various gasket materials to extreme condition science, the optimization of the gasket performance within specific experimental constrains remains one of the most important challenges. The search for the optimal gasket material is always a compromise among the strength, x-ray transparency, chemical inertness, penetrability for gaseous pressure transmitting media (PTM), simplicity for machining, costs and other factors of the material. This study is focused on improvement of gasket material performance for x-ray diffraction applications at conventional and ultra-high pressure conditions.

Ultra-high pressure experiments (those beyond 1 Mbar) always require more elaborative preparation and, at times, unconventional diamond anvils and the matching gasket designs. It was shown that the double-stage DACs (ds-DACs) with the secondary anvils, which enable the pressure multiplication due to a stepwise decrease of the size of a diamond tip, allow one to achieve pressures as high as 1 TPa (Dubrovinskaia *et al.*, 2016). Following the similar idea, the so called toroidal-DACs (t-DACs) were introduced (Dewaele *et al.*, 2018, Jenei *et al.*, 2018, McMahon, 2018). Culets of their anvils are modified using focused ion beam (FIB). The resulting toroidal profile features a central tip with an effectively reduced anvil size of about 15-25 µm in diameter and a few microns (<5 µm) in height. Depending on the shape of the t-DAC tips, the size of the sample chamber is ~3-7 µm in depth and a ~10-15 µm in diameter. Experiments using t-DACs are usually conducted in uniaxial geometry (**Fig. 1c**). Pressures of a few Mbar (beyond 600 GPa) have been achieved in t-DACs experiments. Ultra-high pressure experiments set the ultimate requirement for x-ray diffraction in terms of S/N, as the volume of the sample chamber and the amount of sample surrounded by gasket material are significantly reduced. As we show below, introducing mechanically strong amorphous metal material with optimized yield strength and ductility will be of great benefit for a number of x-ray diffraction experiments, but in particular it will benefit the studies at pressures exceeding megabar.

Amorphous metals (metallic glasses) are non-crystalline materials with a glass-like structure. They do not produce a background in the form of sharp diffraction peaks as crystalline materials do, and still possess high strength. Their properties highlight them as potential candidates for very efficient gasket materials. Gaskets made of metallic glass with the chemical composition $Pd_{40}Ni_{40}P_{20}$ were successfully used in static high pressure XRD experiments beyond 1 Mbar (He *et al.*, 2003). Another amorphous metal, $Fe_{0.79}Si_{0.07}B_{0.14}$, with a significantly lower average number of electrons per atom, was recently used in dynamic high-pressure experiments and showed sufficient stability above 1 Mbar and temperatures >1000 K (Mendez *et al.*, 2020). In the present work, we extend the test of $Fe_{0.79}Si_{0.07}B_{0.14}$ metallic glass performance to several challenging high-pressure experiments including single crystal x-ray diffraction (scXRD) and powder x-ray diffraction (pXRD), involving conventional DACs and t-DACs.

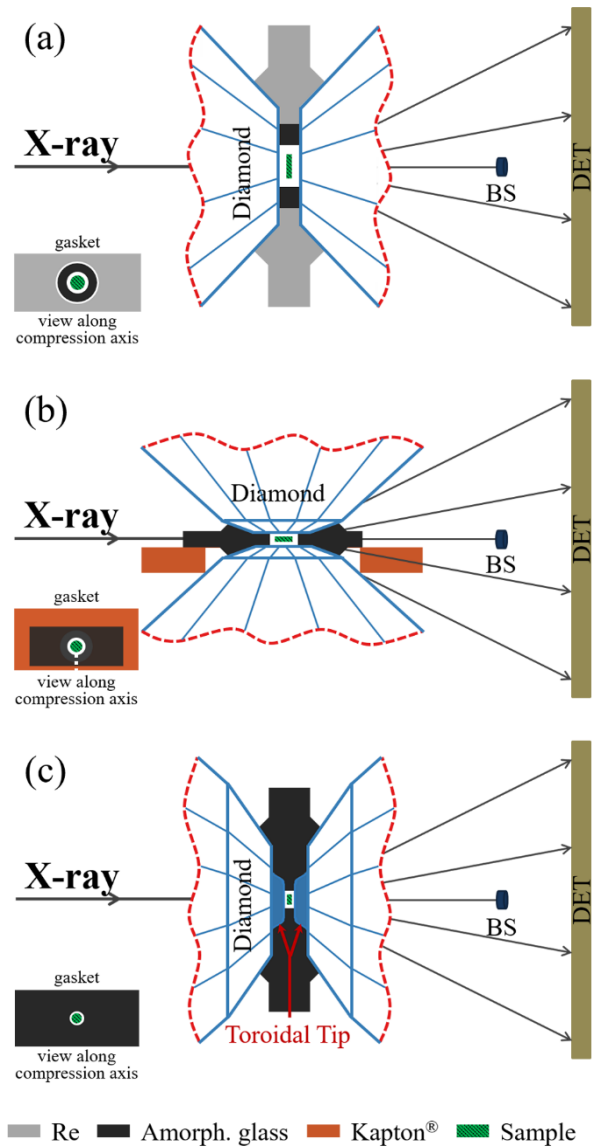

**Figure 1** Schematic drawings of high-pressure XRD experiments in DACs with different geometries. (a) Uniaxial geometry adopted for FeBO$_3$ scXRD with Ne as PTM and for FeH pXRD with H$_2$ as PTM; (b) Radial geometry employed for laser heated GeO$_2$ pXRD in a bevel DAC without PTM; (c) Uniaxial geometry for t-DAC experiment (ultra-high pressure pXRD experiments on Fe with MgO and Au with Ne). Inserts in the lower left corner indicate the gasket assembly as viewed along the DAC compression axis: the sample chamber is shown as a white circle; the white dashed line shown in the insert (b) indicates a slot in the Kapton® foil, making it easier to remove Kapton® after the gasket fixation by the diamond anvils. The materials of the gasket assembly and the sample are colour coded as shown at the bottom: light grey, dark grey, orange and green correspond to Re, Fe$_{0.79}$Si$_{0.07}$B$_{0.14}$ metallic glass, Kapton® and sample, respectively. BS is the beam stop and DET indicates the positions of the detector.

## 2. Experimental methods

Metallic glass with the chemical composition of $Fe_{0.79}Si_{0.07}B_{0.14}$ was used to make gaskets or gasket insets for conventional metallic gaskets. Amorphous metal strips were cut from a melt-spun ribbon with a thickness of 23-25 µm produced in Institute of Physics, P.J. Šafárik University. The chemical composition of the metallic glass was confirmed by the electron microprobe analysis conducted at the Bayerisches Geoinstitut (BGI, Bayreuth, Germany), providing a well consistent result of $Fe_{0.78(6)}Si_{0.06(2)}B_{0.15(7)}$. Mechanical properties of $Fe_{0.79}Si_{0.07}B_{0.14}$ should be remarkably close to those reported elsewhere (Luborsky & Walter, 1980, Hagiwara *et al.*, 1982, Nagarajan *et al.*, 1988, Sypien & Kusinski, 2006, Cadogan *et al.*, 2013) for the Fe-Si-B amorphous alloys of similar composition.

In order to produce sample chambers in the gasket material, we used an electric discharge drilling machine (EDM) and a nanosecond pulse Excimer laser drilling machine operating at 193 nm (Optec™ LaserShot Master). We noticed that the laser drilling may potentially cause recrystallization of glass at the circumference of the sample chamber hole, depending on drilling parameters. That was never the case for the EDM drilling. The recrystallization can be reduced or avoided by optimizing laser operation parameters, e.g. drilling at a low repetition rate (25-50Hz, 6-7mJ per pulse). Despite a minor presence of recrystallized gasket material, experiments with t-DACs showed that such contribution to the diffraction signal is indeed negligible.

All high pressure XRD experiments in this work were performed at the P02.2 (ECB) beamline at PETRA-III, DESY, Hamburg, Germany (Liermann *et al.*, 2015). The instrumental parameters were varying according to experimental conditions as discussed below. The data were collected using a Perkin Elmer XRD 1621 flat panel detector. For the purpose of the data processing, we used various software packages including CrysAlis Pro (Rigaku) (Rigaku, 2015), Olex2 (Dolomanov *et al.*, 2009) in combination with SHELXT (Sheldrick, 2015a, b, 2008), JANA2006 (Petříček *et al.*, 2014), GSAS-II (Toby & Von Dreele, 2013), and DIOPTAS (Prescher & Prakapenka, 2015).

The DAC experiments are divided into several cases (**Table 1**). In case A, a conventional Re gasket with an amorphous metal insert was used in DAC experiments employing uniaxial diffraction geometry (**Fig. 1a**). Here we refer to the examples of single crystal x-ray diffraction (scXRD, case A1) and powder x-ray diffraction (pXRD, case A2). In case B, pXRD experiments were conducted in radial geometry in a DAC supplied with a metallic glass gasket placed on a piece of Kapton® foil (**Fig. 1b**). Case C describes two applications of the metallic glass gaskets in t-DACs for ultra-high pressure pXRD experiments in uniaxial geometry (**Fig. 1c**). In case C1 we compressed Fe together with MgO, while case C2 describes the compression of Au together with Ne. **Table 1** summarizes the information of the experiments including instrumental parameters. The details of the individual experiments and assembly procedures are discussed below.

**Table 1** Summary of the experimental information.

| Case | Sample, PTM | Diamond anvil type | X-ray beam: wavelength, dimension* | Gasket dimension** | Maximum sample pressure, temperature |
|---|---|---|---|---|---|
| **A1** | FeBO$_3$ (single crystal), Ne | Boehler-Almax, ⌀150/300 µm, 8° bevel | 0.2898 Å, 8 × 3 µm$^2$ | ⌀$_{Re}$/⌀$_A$ = 120/75 µm, $t$ = 20 µm | ~1.3 Mbar, 300 K |
| **A2** | FeH (sub-micron grain powder), H$_2$ | Boehler-Almax, ⌀300 µm | 0.4845 Å, 2 x 2 µm$^2$ | ⌀$_{Re}$/⌀$_A$ = 200/100 µm, $t$ = 30 µm | ~0.4 Mbar, 300 K |
| **B** | GeO$_2$ (sub-micron grain powder), no PTM | Standard design, ⌀150/300 µm, 8° bevel | 0.2898 Å, 8 × 3 µm$^2$ | ⌀$_K$/⌀$_A$ = 400/40 µm, $t$ = 25 µm | ~1.0 Mbar, below 2000 K |
| **C1** | Fe (sub-micron grain powder), MgO | t-DAC, tip ⌀20 µm, tip height ~4.3 µm | 0.4834 Å, 1.5 × 0.9 µm$^2$ | ⌀$_A$ = 10 µm, $t$ = 4 µm | ~3.0 Mbar, 300 K |
| **C2** | Au (sub-micron grain powder), Ne | t-DAC, tip ⌀20 µm, tip height ~4.3 µm | 0.4839 Å, 1.3 × 1.0 µm$^2$ | ⌀$_A$ = 10 µm, $t$ = 4-5 µm | ~2.5 Mbar, 300K |

*Beam size at the focal spot corresponds to H x V at FWHM.

**⌀$_{Re}$, ⌀$_A$, and ⌀$_K$ correspond to the diameters of holes inside the Re, metallic glass and Kapton® pieces, respectively.

$t$ - gasket thickness before sample loading.

**Case A: conventional Re gasket featuring an amorphous metal insert for scXRD and pXRD experiments in uniaxial geometry**

The amorphous metal gasket for the scXRD experiment (case A1) was prepared as follows. Firstly, a 250 µm thick Re foil was pre-indented with a pair of diamonds with the culet size of 150 µm/300 µm bevel at 8° (Boehler-Almax design) to a thickness of ~26-28 µm (slightly thicker than the metallic glass ribbon). Then, a hole of ~110-120 µm in diameter was drilled in the center of the indentation using EDM. A disc of a metallic glass was inserted into the hole and fixed within the primary Re gasket by a gentle compression of the latter between the anvils. Next, a central hole of ~75 µm in diameter (serving as the sample chamber) was drilled in the Fe$_{0.79}$Si$_{0.07}$B$_{0.14}$ insert using EDM. The preparation process described above is somewhat flexible and the sequence of the preparation steps can be adjusted. For example, one can first drill a hole in a metallic glass disc and then insert it into the hole of the Re gasket as was shown in the previous study by Mendez et al. (Mendez *et al.*, 2020).

The pXRD experiment (case A2) was conducted in the pressure range up to 39 GPa by using a pair of 300 µm culet diamond anvils. A Re gasket was first indented to the thickness of 50-55 µm and then drilled a 250 µm diameter hole with EDM. In addition, we cut two disks of amorphous metal (23-25 µm in thick) with similar outer diameter and loaded them as a stack, one on top of the other, into the Re gasket hole installed on a diamond anvil. The discs were fixed to the primary Re gasket by a gentle compression. Finally, the hole forming the sample chamber with a 150 µm inner diameter was drilled with EDM inside of the $Fe_{0.79}Si_{0.07}B_{0.14}$.

**Case B: metallic glass gasket placed on a piece of Kapton® for pXRD experiment in radial geometry**

**Fig. 1b** illustrates the schematic diagram of the radial pXRD experiment where the entire gasket was made from the amorphous metal. The thickness of the as-produced metallic glass ribbon is thin enough (~23-25 µm), so the process of the gasket indention can be saved. To enable precise positioning the tiny sample chamber on the tip of a diamond anvil we supplied the gasket with a support of a 125 µm thick Kapton® (DuPont™) foil. The Kapton® strip with a width of 3 mm was drilled by the Excimer laser drilling machine producing a hole of ~400 µm in diameter. The hole served as a mounting aperture centered with the diamond of 150/300 µm 8° bevel culet. Next, a metallic glass strip of a rectangular shape (width below 1.4 mm) was cut from the metal glass ribbon using scissors and secured with instant glue above the hole in the Kapton® foil. The dimensions of amorphous glass strip can be potentially reduced without compromising mechanical stability of the sample chamber keeping the sample at desired pressures. Last, a hole (concentric to the hole in the Kapton® foil) with a diameter of 40 µm corresponding to the sample chamber was drilled in the metallic glass using Excimer laser drilling machine.

Kapton® piece carrying amorphous gasket can be easily attached to the diamond. During compression, the Kapton® strip may bend or tilt. As a final step of the optimization, it can be fully removed from the cell assembly after the sample was loaded and compressed. This will reduce Kapton® contribution to diffraction signal. In order to simplify the removal procedure, a slit can be cut into the Kapton® strip as illustrated in the insert of **Fig. 1b**.

A modified BX90 cell (Kantor *et al.*, 2012) dedicated for radial high-pressure XRD was employed in this test. The modification concept of BX90 comes from Lowell Miyagi. By modifying the piston and cylinder parts of the original BX90 design, two apertures perpendicular to the compression axis were expanded. This modification enabled data collection in radial geometry within a diffraction scattering cone of ~54.5°. The cells were produced by Extreme Conditions Science Infrastructure of PETRA-III, DESY.

**Case C: metallic glass gasket in t-DACs for pXRD experiment in uniaxial geometry**

In our test we used a t-DAC in order to demonstrate the capability of the material under pressures of a few Mbar. In case C1 we used Fe together with MgO, while in case C2 Au was loaded together with Ne PTM. The toroidal profiles of the diamond culets were prepared by using a focused ion beam instrument (FIB, SCIOS, FEI Thermofisher) installed at DESY's NanoLab (Stierle *et al.*, 2016) within the framework of the BMBF project No. 05K13WC3 (PI. Natalia Dubrovinskaia). The milling bitmap is shown in **Fig. A1** along with the image of the resulting profile of the diamond culet, which is similar to the previous design suggested by Dewaele et al. (Dewaele *et al.*, 2018). After 4 hours of milling with the ion beam of 15 nA under 30 kV acceleration voltage, we can produce a single toroidal anvil with a conical shape tip of ~4.3 µm in height and ~20 µm in diameter. A piece of ~25 µm thick metallic glass ribbon with a hole of 40 µm was directly placed between the toroidal anvils for the first indention. During the indentation, the diamond anvils were brought together and visible light interferometry was used to measure the distance between the toroidal tips. At the moment of the hole collapse, if the distance was still larger than the target thickness of 4-5 µm, a new hole was made and the process of indentation was repeated. The process was stopped until the target thickness was achieved. The final sample chamber of ~10 µm in diameter was drilled with several shots of Excimer laser.

## 3. Results and discussions

### 3.1. Single crystal x-ray diffraction of FeBO$_3$ beyond 1 Mbar in a DAC with Ne PTM

The DAC assembly was prepared as described above for the case A1, and the experiment design is illustrated in **Fig. 1a**. A single crystal of FeBO$_3$ (Kotrbová *et al.*, 1985) was loaded into the sample chamber together with Ne as PTM and pressure marker (Fei *et al.*, 2007). The sample was compressed to ~1.3 Mbar in this experiment. The scXRD data were collected during a rotation of the sample within ±32° around the ω axis (perpendicular to the x-ray beam) with a step size of 0.5°. Additional technical information is summarized in **Table 1**.

The collected XRD data is of excellent quality despite the megabar conditions and the occurrence of a spin-state transition at ~50-54 GPa (Gavriliuk *et al.*, 2002, Vasiukov *et al.*, 2017). **Fig. 2a** shows a section of the 2D diffraction pattern at 103.3(1) GPa with reflections of FeBO$_3$ (single spots), Ne PTM, and diamond anvil Bragg reflections (masked in red) labelled. The contribution of gasket material to the detected signal is negligible despite the relatively large x-ray beam size used in this experiment (see **Table 1**). If the gasket would have been made entirely of Re or W, the diffraction lines of these strongly scattering materials would overlap with the diffraction spots of the single crystal substantially reducing the quality of the data. Here we focus on a single pressure point of 103.3(1) GPa as the full data set will be published elsewhere.

Indexing of the scXRD using CrysAlis Pro software confirmed the trigonal space group $R\bar{3}c$ for FeBO$_3$ at 103.3(1) GPa. The full crystallographic information is provided in **Table 2**. Supplementary information contains the corresponding CIF file. **Fig. 2b** shows the integrated 1D diffraction pattern and the Le-Bail fit result performed by the GSAS-II software package. The Le-Bail refinement converged to the unit cell parameters $a$ = 4.2555(8) Å, $c$ = 11.796(7) Å, which is in good agreement with the structure solution and refinement results based on scXRD (see **Table 2**).

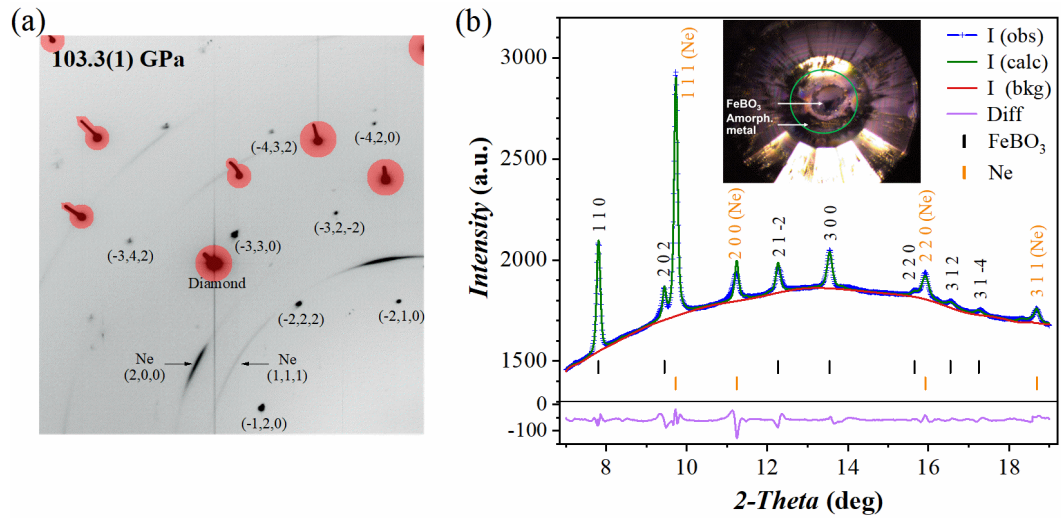

**Figure 2** Diffraction patterns of FeBO$_3$ single crystal in a DAC with Ne PTM at 103.3(1) GPa. (a) A representative section of a 2D diffraction pattern. The Bragg peaks from diamond anvils are shaded in red; representative peaks of FeBO$_3$ and Ne are indexed. (b) 1D diffraction pattern. The inset shows a microscope photograph of the sample chamber indicating the position of the FeBO$_3$ single crystal and the inner edge of Re gasket (green circle) which is also the outer edge of the amorphous metal insert; for the scale, the diamond culet diameter is 150 μm. No strong scattering from the gasket material was observed, although the x-ray beam was focused to 8 x 3 μm$^2$.

**Table 2** Crystallographic information for FeBO$_3$ single crystal at 103.3(1) GPa.

| | |
|---|---|
| **Crystal information:** | |
| Exp. wavelength (Å) | 0.2898 |
| Space group (No.) | $R\bar{3}c$, 167 |
| Z | 6 |
| $a$ (Å), $c$ (Å) | 4.2510(2), 11.772(12) |
| $V$ (Å$^3$) | 184.2(2) |
| **Refinement details:** | |
| $R(F^2)$; $I > 2\sigma(I)$ | 5.8 % |
| $wR(F^2)$; $I > 2\sigma(I)$ | 12.2 % |
| *Completeness (%)* | 24.6 |
| $N_{ref}/N_{par}$ | 42/5 |
| HKL statistics | $-8 \leq H \leq 8$; $-7 \leq K \leq 6$; $-4 \leq L \leq 5$ |

**Structural parameters:**

| Atom | Site Sym. | Atomic coordinates ($x, y, z$) | $U_{iso}$ |
|---|---|---|---|
| Fe | 6b | 0, 0, 0 | 0.0037(7) |
| O | 18e | 0.3094(10), 0, 0.25 | 0.0050(10) |
| B | 6a | 0, 0, 0.25 | 0.0110(30) |

$R(F^2)$, $wR(F^2)$ – crystallographic R-factors, I ~ F$^2$ parameter corresponds to intensity, while F is the scattering factor

$N_{ref}$, $N_{Par}$ – number of unique reflections (merged) used in analysis, number of refined parameters, respectively

$U_{iso}$ – isotropic thermal displacement parameter

### 3.2. Powder x-ray diffraction of FeH at 39 GPa in a DAC with H$_2$ PTM

We expanded our investigation of Fe$_{0.79}$Si$_{0.07}$B$_{0.14}$ metallic glass performance further and loaded a DAC with a lighter and much more chemically active PTM of H$_2$. The sample chamber was prepared as described above (see case A2). The experimental geometry is shown in **Fig. 1a**. We loaded iron ($^{57}$Fe enriched to 91%, purity 96.28% trace metals basis, sold by CHEMGAS®), ruby as the pressure marker and H$_2$ as PTM into the sample chamber. **Fig. 3** shows the microscope photographs of the sample chamber compressed to around 2-3 GPa (**Fig. 3a**) and at ~39 GPa after equilibration for three weeks (**Fig. 3b**). During the timespan of three weeks the sample chamber did not collapse indicating low permeability of H$_2$ through the gasket material.

We performed XRD mapping over the entire sample chamber (see **Table 1** for technical details). The representative 2D and 1D diffraction patterns are shown in **Fig. 4**. It is well known that Fe starts to react with H$_2$ at very low pressures forming iron hydride FeH (Hirao *et al.*, 2004). This reaction was confirmed in our pXRD data. The unit cell parameters of FeH were determined by Le-Bail refinement using GSAS-II resulting in: $a$ = 2.556(1) Å, $c$ = 8.33(1) Å (space group *P6$_3$/mmc*). It is almost impossible to detect H$_2$ in the diffraction data (as illustrated in **Fig. 4a**) because of several effects: the low scattering power of $H_2$, large scattering contrast between Fe and H (short accumulation times) and the well-known effect of $H_2$ recrystallization in the form of larger single crystal-like grains rather than a fine-grain powder. Additional analysis of Raman spectra measured separately confirmed presence of $H_2$ vibron (Goncharov *et al.*, 2011). At the boundary of the sample chamber, the XRD map did not indicate formation of any Fe, B or Si bearing polycrystalline compounds (e.g. hydrogen containing) or recrystallized metallic glass. Our observations confirm that $H_2$ diffuses much less through $Fe_{0.79}Si_{0.07}B_{0.14}$ in comparison to W, Re or steel, which will quickly react with hydrogen. It is known that reaction between $H_2$ and Re may lead to a collapse of the pressure chamber with time. The performance of this amorphous metal with other highly permeable gaseous PTMs, e.g. He, should be investigated additionally, but the results on $H_2$ and Ne are very promising.

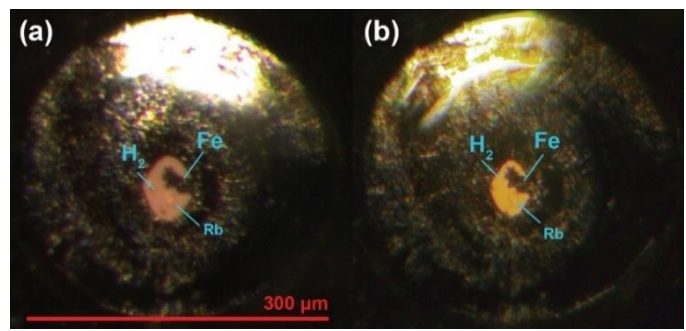

**Figure 3** Microscope photographs of the sample chamber with $^{57}$Fe and ruby in $H_2$ PTM: (a) at 2-3 GPa; (b) at 39 GPa after 3 weeks of equilibration.

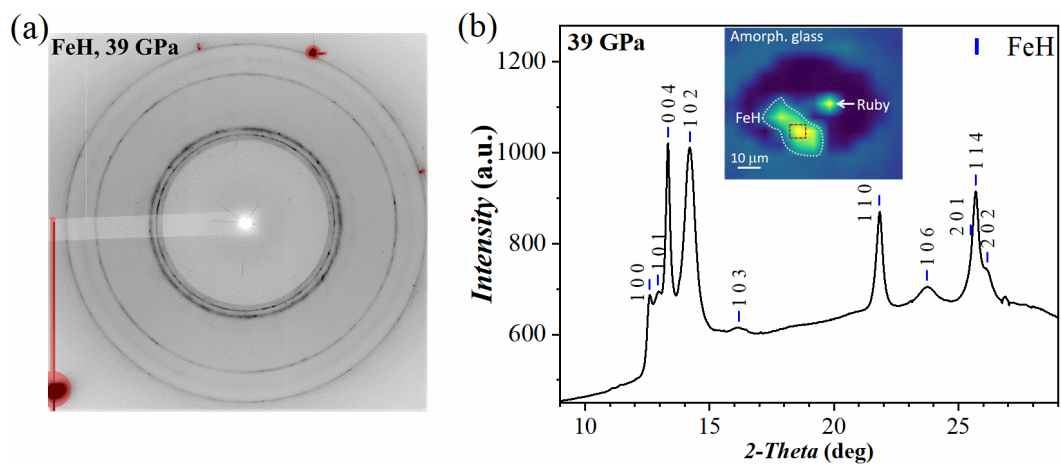

**Figure 4** Diffraction pattern of FeH at 39 GPa with $H_2$ as PTM loaded into the amorphous metal insert in a Re gasket. (a) 2D diffraction pattern with no significant contribution from gasket material. Hydrogen signal is barely visible due to the low scattering power. Diamond Bragg peaks are masked with red markers. (b) The corresponding integrated 1D diffraction pattern including the index of FeH peaks. The insert shows the reconstruction of phase distribution obtained from the analysis of 2D pXRD map produced by sample scanning. We indicate the position of FeH and ruby in the sample

chamber surrounded by amorphous metal. The boundary of the amorphous metallic material can be clearly seen.

### 3.3. Radial powder x-ray diffraction of laser heating GeO$_2$ at > 1 Mbar in a DAC with no PTM

The DAC assembly was prepared as described above for the case B. The experiment setup is illustrated in **Fig. 1b**, and we refer to **Table 1** for additional experimental details. We loaded a mixture of GeO$_2$ powder (Sigma Aldrich 483702, ≥99.99% trace metals basis) and small amount of Pt powder, serving as laser light absorber, into the sample chamber. The sample was compressed to >80 GPa and then laser heated at above 1000 K at two facilities (at BGI with pulsed laser, and at P02.2 with continuous laser). The sample was further compressed to about 90 GPa. The pressure was estimated from the diamond-Raman shift (Dubrovinskaia *et al.*, 2010). Repeated laser heating of the sample even up to ~2000 K in the vicinity of 90 GPa did not cause any detectable damage to neither the metallic glass gasket nor to the diamonds. After heating at ~90 GPa, the sample was further compressed at ambient temperature to pressures exceeding one megabar.

**Fig. 5** shows the 2D and 1D pXRD patterns collected at 103(1) GPa in radial geometry. They are dominated by a few intense but relatively broad peaks originating from the amorphous metal and several sharp diffraction lines contributed by the pyrite-structured polymorph of GeO$_2$. The amount of Pt in the cell was rather small, so that Pt peaks are not visible in the diffraction pattern. Although the scattering from the metallic glass gasket in radial geometry is omnipresent (**Fig. 5b**), it does not prevent a reliable data analysis when it manifests as a smooth background. Indeed, it is very easy to fit the background profile and subtract from the raw 1D diffraction pattern (as shown in **Fig. 5b**), which offers one of the biggest advantages in comparison to the conventional crystalline gasket materials (e.g. Be). The data were processed using GSAS-II software package. The unit cell parameter of GeO$_2$ ($Pa\bar{3}$ space group) was determined to be $a$ = 4.3591(3) Å. As we mentioned earlier, the contribution of amorphous gasket scattering signal can be further reduced by optimization of the amorphous gasket dimensions. It is also important to mention that we did not observe any additional contribution to the signal from potentially recrystallized glass after repetitive laser heating the GeO$_2$ sample to 1000-2000 K, although we produced the pyrite-GeO$_2$ in the entire area of the sample chamber.

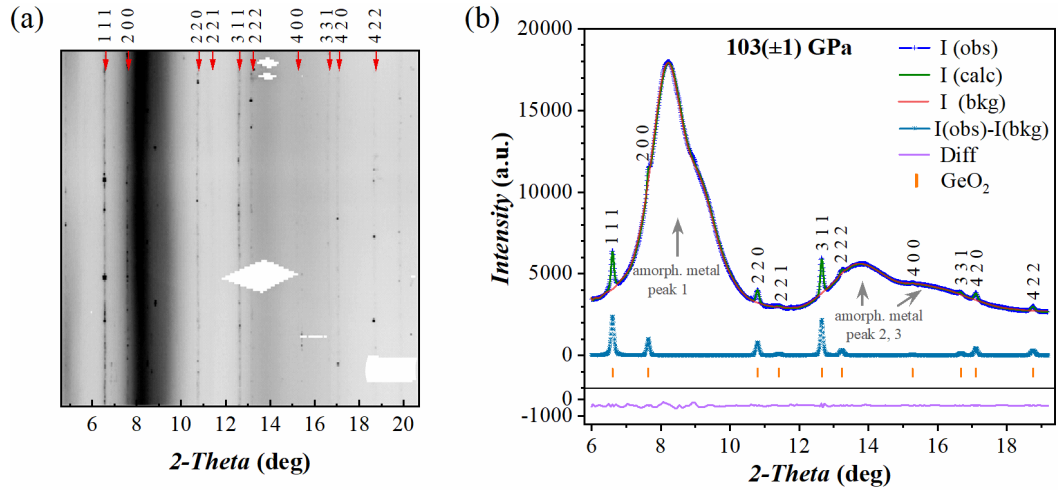

**Figure 5** Radial diffraction patterns of $Pa\bar{3}$ GeO$_2$ polymorph at 103(1) GPa. (a) Unwrapped 2D diffraction pattern. Diffraction lines of GeO$_2$ are indexed. The narrow and straight diffraction lines indicate small grains and low amount of strain in the sample. Diamond peaks were masked prior to 1D pattern integration. (b) The matching 1D diffraction pattern and the corresponding GSAS-II fit. The raw intensities - I(obs) are shown using the dark blue line. The light blue line corresponds to the raw data with the background intensity - I(bkg) subtracted. Tick marks indicate positions of the diffraction lines from GeO$_2$ at 103(1) GPa. I(calc) and Diff correspond to the calculated intensities (the green line) and the residuals of the Le-Bail fit (the purple line), respectively.

### 3.4. Powder x-ray diffraction of Fe + MgO at multi-Mbar pressure in a t-DAC

The DAC assembly was prepared as previously described for the case C. See also **Fig. 1c** and **Table 1** for additional experimental details. We will start our discussion with case C1 focusing on the compression of metallic Fe together with MgO used as PTM.

**Fig. 6a** shows a microscope photograph of the primary diamond anvil with the toroidal profile. A 3D image of the tip profile produced by the atomic force microscope (AFM, Dimension Icon, Bruker) at the DESY Nanolab is shown in **Fig. 6c**. The pressure chamber was loaded with a piece of iron powder (~2-5 µm in diameter, Alfa Aesar 00170, 99.9+%) as sample material, and nano crystalline powder of MgO (Sigma Aldrich 549649, particle size ≤50 nm) as pressure marker and PTM. Small grain size powders were selected in order to achieve a more uniform signal distribution along the Debye-Scherrer ring. While our Fe powder had larger grains in comparison to MgO, the transition of iron to hexagonal closed packed (hcp) phase led the recrystallized material to a finer grain powder resulting in a more homogeneous distribution of intensity along diffraction rings.

The sample chamber containing Fe and MgO was compressed with a step size of ~30 GPa in the t-DAC up to 300 (5) GPa. Pressure was determined using the MgO equation of state (EoS) of Jacobsen et al. (Jacobsen *et al.*, 2008) with $K_0$ = 159.6 GPa and $K'$ = 3.74 (where $K_0$ and $K'$ are the bulk modulus and it's pressure derivative at 1 bar). If the MgO EoS of Zha et al. with parameters of $K_0 = 160.2$ GPa and $K' = 4.03$ (Zha *et al.*, 2000) is used, the highest stress condition experienced by MgO equates to 340(5) GPa. Here we just show a snapshot from a larger work which will be published elsewhere.

At each compression step, the pressure chamber and its vicinity were scanned and 2D x-ray transmission maps were produced (**Figs. 7a-c**). The data was used to monitor the integrity of the pressure chamber and deformation of the diamond anvils. The 2D XRD patterns were also obtained in the scanning regime and the data corresponding to 280(5) GPa is shown in **Fig. A2**. The corresponding integrated 1D XRD pattern is shown in **Fig. 7d** together with the representative 2D pattern. As shown in **Fig. A2**, few reflections of *hcp*-Fe and two weaker reflections of MgO are recorded. However, our observations indicate a finite contribution of preferred orientation of MgO particles developed during the compression. The data are of high quality even at the highest-pressure conditions: signal intensity is strong, and the S/N is reasonably improved in comparison to conventional ultra-high pressure experiments involving polycrystalline Re or W gaskets. The lattice parameters were found to be $a = 2.114(2)$ Å, $c = 3.378(3)$ Å for *hcp*-Fe and $a = 3.459(1)$ Å for MgO.

As described above, the experiment was conducted under nonhydrostatic conditions. Compression in t-DAC generates enormous stresses and strains which are partially responsible for the visible peak broadening (**Fig. 7d**). The nonhydrostatic conditions may also be a reason for a large difference in pressures calculated using either MgO EoSes (Jacobsen *et al.*, 2008, Zha *et al.*, 2000) or Fe EoSes (Dewaele *et al.*, 2006). Considering the data shown in **Fig. 7d,** the Fe EoSes estimate the pressure to be ~338 - 347 GPa (according to Vinet and the 3$^{rd}$ order Birch Murnaghan EoS of Dewaele et al., respectively). These values are larger than those determined using MgO EoSes (Jacobsen *et al.*, 2008, Zha *et al.*, 2000). Here our observations should stimulate the community attention with respect to the challenges of multiphase compression and pressure determination at non thermally equilibrated conditions in conventional DACs in general, e.g. (Glazyrin *et al.*, 2016), and in t-DACs in particular.

In our t-DAC experiments we used x-ray energy at 25.6 keV. The accessible Q-range in t-DAC at this energy range is quite limited. It may present a challenge in interpretation of the XRD data for a great number of inorganic materials, including simple solids under pressures exceeding 300 GPa. Access to x-ray sources with higher energy (allowing larger Q-range), higher flux (allowing stronger signal from smaller scattering volume), as well as smaller beam at the focal spot (allowing more precise mapping of stress and strain state of the sample and suppressing 'parasitic' scattering from gasket material) would be beneficial to t-DACs experiments at multi-megabar pressures. Our experiments are going along the community requests towards the newer generation of experiments which could be much improved with x-ray sources such as DLSRs and x-ray FELs. Both are considered the next step for exploration path of extreme conditions, and considering that the current generation of XFELs are impressing machines with superior time domain resolution, but operating at energies up to 25 keV, it becomes clear that only a combination of XFEL and SR facilities will enable the thorough characterization of multi-megabar samples compressed in t-DACs, especially in situations where single crystal diffraction analysis has to be involved (e.g. 'cook-and-look' experiments and etc.).

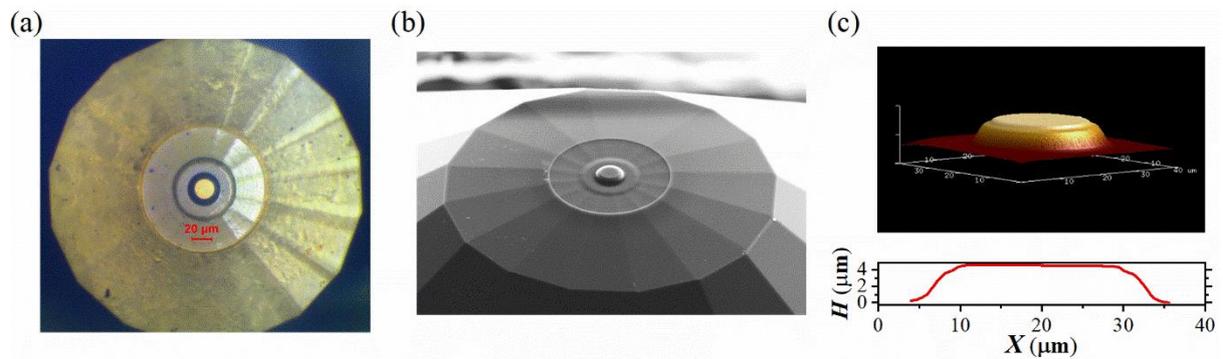

**Figure 6** The toroidal diamond anvil fabricated from a standard diamond anvil (Almax-diamonds) with the culet size of 40/300 μm, bevel angle of 8°. (a) Microscope photograph of the toroidal diamond anvil. (b) Photograph taken by electron beam in the FIB chamber with the sample stage inclination of 52°. (c) The tip of the toroidal anvil recorded by AFM.

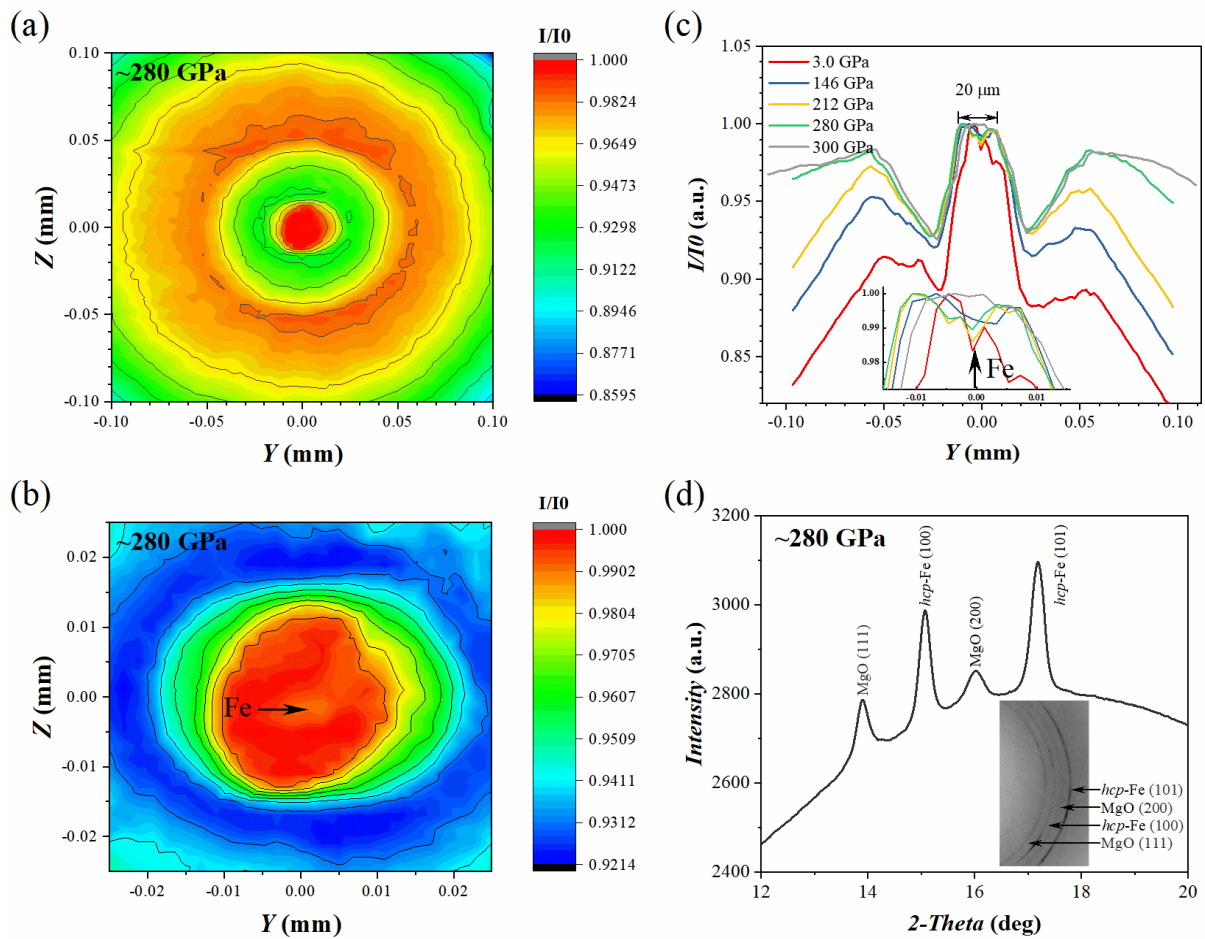

**Figure 7** X-ray transmission maps (1D and 2D) across the anvils of the t-DAC and XRD patterns collected from the Fe + MgO sample. (a) The mapped area of 200 x 200 μm² (step size of 4 μm)

displays the transmission profile of the entire primary anvil culet which is increased to ~105-110 μm after manufacturing of toroidal tip; (b) The mapped area of 50 x 50 μm$^2$ offers a detailed view (step size of 2 μm). The position of iron is indicated by the arrow and confirmed by 2D pXRD mapping; (c) 1D toroidal anvil x-ray transmission profiles at different pressures; (d) The integrated 1D diffraction pattern of Fe and MgO sample at ~280(5) GPa. The inset shows a part of a 2D diffraction image. The pressures here were calculated with Jacobsen et al. MgO EoS (Jacobsen *et al.*, 2008). I/I$_0$ corresponds to the x-ray intensity ratio of the transmitted and the incident beams.

**3.5. Powder x-ray diffraction of Au + Ne at multi-Mbar pressure in a t-DAC**

If we consider various aspects of quickly developing t-DAC techniques, we quickly come to an understanding that quasihydrostatic pressure conditions as well as the precision of pressure determination are of great importance, but the topics are not well explored. Here, we extended our tests of t-DAC and present the case C2, where gold particles were loaded into a sample chamber of a t-DAC together with Ne as PTM, which are often used as pressure sensor materials in methodological studies. Additional details can be found in **Table 1**. **Fig. 8a** shows a microscope photograph of the t-DAC cell pre-loaded to ~65 GPa confirmed by diamond Raman peak measured at the tip of the toroid anvil. Great mechanical properties of gasket material allowed us to compress Ne and Au clamped between the toroidal tips to above 2 Mbar. **Fig. 8b** and **c** present the 2D and 1D diffraction patterns of Au and Ne at the highest pressure we achieved. Exceptionally simple background of the diffraction patterns (**Fig. 8b, c**) allowed determination of the unit cell volumes of Au ($V_{Au}$) and Ne ($V_{Ne}$) as the function of pressure. In this t-DAC experiment, we investigated that the pressure estimation is strongly dependent on the choice of EoS. In **Fig. 8c**, we show the pressure results calculated via different Ne EoSes (Dorfman *et al.*, 2012, Fei *et al.*, 2007). We also observed a little peak position shift of Au between the individual peak profile fitting and the LeBail fitting results. As shown**,** the black and red ticks labelled in **Fig. 8c** correspond to Au *hkl* reflections by LeBail fit and individual peak profile fitting respectively. A small offset of Au (200) could be attributed to a small deviatoric stress. It is similar to the behavior of Ne (200) at low and moderate pressures.

The small number of visible diffraction peaks and their low intensities prevent us from performing a full deviatoric stress analysis for neither Au nor Ne and, thus characterize the magnitude of non-hydrostatic stresses and strains. But it is equally important to show a clear experimental evidence indicating deviatoric stress presence in t-DAC experiments, emphasizing a complicated picture of multiphase compression in the latter.

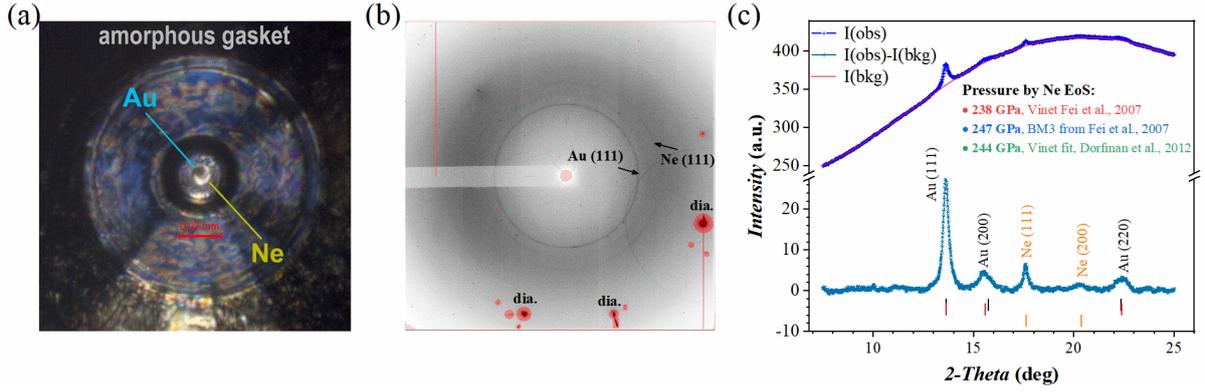

**Figure 8** (a) Microscope photograph of t-DAC assembly loaded with Au and Ne at ~65 GPa. Darker area inside the sample chamber corresponds to Au powder, while the brighter area is illuminated by back light corresponds to Ne. (b) 2D x-ray diffraction pattern collected at ~247 GPa (pressure determined by BM-3$^{rd}$ EoS of Ne from Fei et al. (Fei *et al.*, 2007)) with indication of (111) peaks of Au and Ne. Strong diamond peaks are indicated separately. (c) 1D diffraction pattern corresponding to (b). With respect to the x-ray background signal on 1D pattern, we observe a continuous variation of intensity attributed to the contribution from Compton scattering of the diamond anvils and a negligible contribution from the amorphous gasket. The raw data - I(obs) is indicated with a dark blue line; the light blue line corresponds to the raw data with background intensity - I(bkg) subtracted. Black/red and orange ticks shown under the 1D pattern indicate peak positions for Au and Ne, respectively. The black ticks correspond to Au *hkl* reflections with LeBail fit, while the red ticks indicate diffraction peak positions of Au fitted individually. The pressure values were estimated by different Ne EoSes. BM3 corresponds to the 3rd order Birch- Murnaghan EoS.

**Fig. 9a** shows the volume of Au ($V_{Au}$) as a function of the volume of Ne ($V_N$) with compressing in a t-DAC. Considering the literature data, in **Fig. 9a** we also present the calculated $V_{Au}$ and $V_{Ne}$ using some of the pressure scale studies reporting simultaneous Au and Ne compression. (Fei *et al.*, 2007, Dorfman *et al.*, 2012) Right and top axes indicate pressure scales calculated from the EoSes with the corresponding labels shown in the figure. Although there is no perfect match, we could imagine better matching results when using the 3$^{rd}$-BM EoS from Fei et al. Small deviations are observed when comparing our data with the literatures, e.g. Dorfmann et al. However, the latter literature data show a strong overlap of the diffraction peak from Au, Re and even NaCl, which could have led to interferences during data analysis. Diffraction patterns shown in **Fig. 8** represent a great example illustrating the advantages of amorphous metallic gasket (with respect to the background and sample signal overlapping issue) over the crystalline gaskets.

We supplement the discussion with **Fig. 9b**, where we present an average pressure difference of $\overline{\Delta P} = \overline{P_{Ne} - P_{Au}}$ as a function of the whole average pressure $\overline{P} = \overline{P_{Ne} + P_{Au}}/2$ calculated from the EoSes shown in **Fig. 9a**. The data indicates that during the compression process in t-DAC loaded with Au and Ne, we got the pressure deviation below 5%. We consider this observation as important, especially given to the tiny dimension of the t-DAC sample chamber and the condition of great stress and strain. Our data inspires cautious optimism that EoSes of Ne and Au measured at lower pressure range can be indeed applied to t-DAC loadings with quasihydrostatic pressure media, and can be reasonably extrapolated to higher pressures (e.g. EoS of Fei et al.).

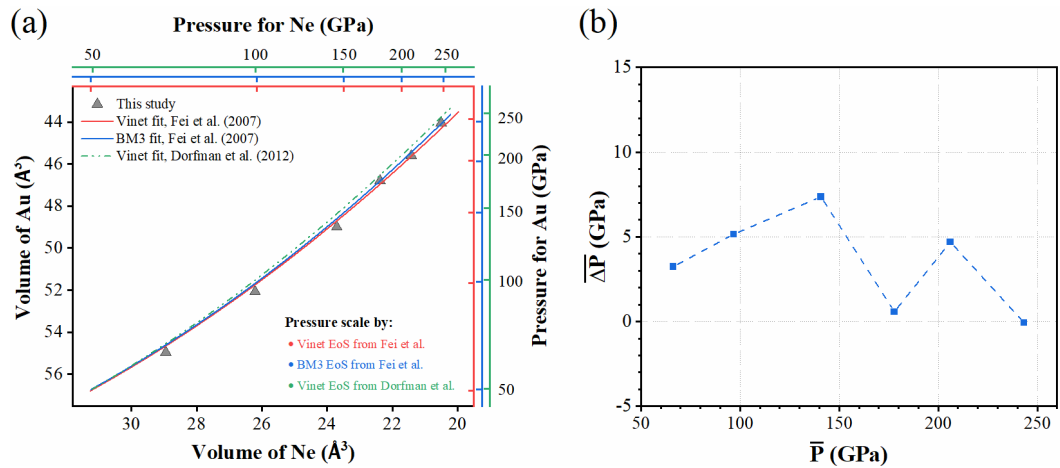

**Figure 9** Data describing Ne + Au compressed in a t- DAC at room temperature in comparison with the literatures (Fei *et al.*, 2007, Dorfman *et al.*, 2012): (a) measured unit cell volumes of Ne and Au are compared with EoSes from the literatures; (b) average pressure deviation between Au and Ne as a function of the average pressure calculated using the indicated EoSes. The former value is calculated as $\overline{\Delta P} = \overline{P_{Ne} - P_{Au}}$, while the latter is defined as $\bar{P} = \overline{P_{Ne} + P_{Au}}/2$.

## 4. Conclusions

In this work we studied the performance of gaskets and gasket inserts made of $Fe_{0.79}Si_{0.07}B_{0.14}$ metallic glass in several high-pressure single crystal and powder x-ray diffraction experiments at pressures ~1 Mbar and above. Although the high pressure/temperature stability of $Fe_{0.79}Si_{0.07}B_{0.14}$ has not been investigated in detail so far, our results show that the material performs well and exhibits sufficient mechanical stability for the DAC assembly even at pressures as high as a few Mbar and even when in contact with sample heated to ~2000 K. The implementation of $Fe_{0.79}Si_{0.07}B_{0.14}$ metallic glass gasket paves a way for improving of the S/N in x-ray diffraction work, which is indeed of great importance for ultra-high pressure experiments in t-DACs and dsDACs aiming for extremes, but also beneficial for experiments at lower pressures conducted with a larger x-ray beams.

Our tests confirmed that $Fe_{0.79}Si_{0.07}B_{0.14}$ can efficiently hold conventional and exotic gaseous PTM, like Ne and $H_2$. Sample laser heating or other high temperature treatment (e.g. resistive heating by Mendez et al. (Mendez *et al.*, 2020)) of the sample chamber during high-pressure experiments may cause partial recrystallization of the amorphous metal material, but we found that the contribution from the recrystallized material is neglectable because of the considerably lower scattering (low Z) in comparison to conventional Re and W gaskets.

We studied the performance of a single Fe-Si-B amorphous metal composition ($Fe_{0.79}Si_{0.07}B_{0.14}$), but the Fe-Si-B phase diagram offers a broad range of compounds. Many of those can be potentially quenched into amorphous state. Therefore, we further refer to (Miettinen *et al.*, 2019) for additional information regarding other possible metallic glass materials within the Fe-Si-B family and their crystallization paths, which, along with $Fe_{0.79}Si_{0.07}B_{0.14}$, could be also potentially used as novel amorphous gasket materials for application in diverse high-pressure high-temperature studies at the 3$^{rd}$ and 4$^{th}$ generation synchrotrons as well as at the x-ray free electron laser facilities.

**Acknowledgements** We thank A. Chumakov (ESRF, EBS, Grenoble, France, Kurchatov Insitute, Moscow, Russia) and G. Smirnov (Centre of Fundamental Research, Kurchatov Insitute, Moscow, Russia) for providing the iron borate single crystal grown by M. Kotrbová and the coworkers. We thank S. Kulkarni (DESY, Hamburg, Germany) for helping us testing the toroidal tip profile measurements using AFM. We acknowledge DESY (Hamburg, Germany), a member of the Helmholtz Association HGF, for the provision of its experimental facilities including Beamline P02.2 (PETRA III) and NanoLab. N.D. and L.D. thank the Federal Ministry of Education and Research, Germany (BMBF, grants No. 05K13WC3 and 05K19WC1) and the Deutsche Forschungsgemeinschaft (DFG projects DU 954-11/1, DU 393-9/2, and DU 393-13/1) for financial support. N.D. thanks the Swedish Government Strategic Research Area in Materials Science on Functional Materials at Linköping University (Faculty Grant SFO-Mat-LiU No. 2009 00971). This work was partially supported by the Slovak Science Grant Agency - project VEGA 1/0406/20.

**Appendix A. Bitmap and fabrication result of the t-DAC and full 2D diffraction pattern of Fe + MgO in a t-DAC**

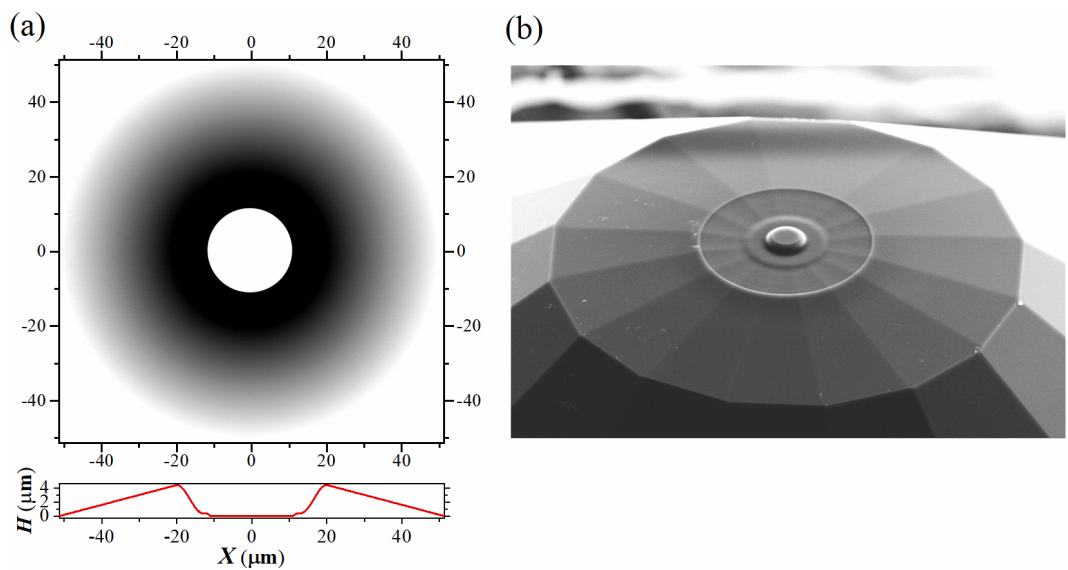

**Fig. A1**  (a) Bitmap employed for the toroidal diamond anvil fabrication. Toroid modification was produced on top of a standard diamond (Almax-diamonds) with the initial culet size of 40/300 μm, bevel angle of 8º. The 1D milling profile is indicated. (b) A photograph made in FIB presenting a view of the milled diamond with an inclination of mounting stage of 52º.

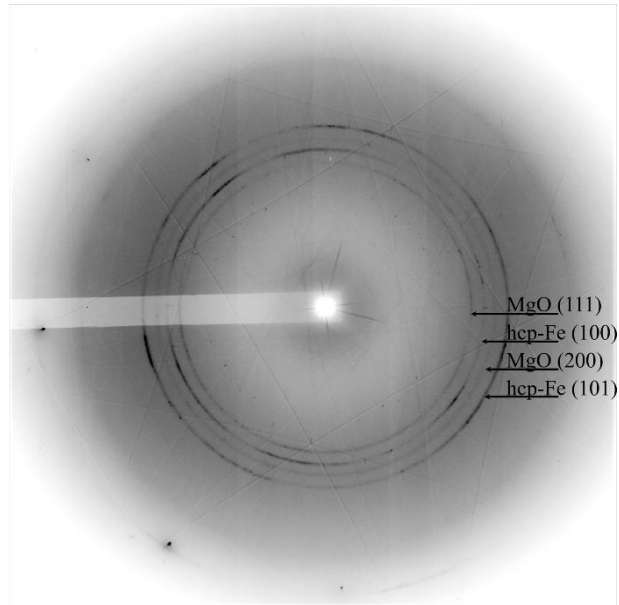

**Fig. A2**  The full 2D diffraction pattern of Fe + MgO collected at ~280(5) GPa (MgO EoS (Jacobsen et al., 2008)) or 313(5) (MgO EoS (Zha et al., 2000)) collected from the t-DAC. The sample chamber was surrounded by $Fe_{0.79}Si_{0.07}B_{0.14}$ metallic glass gasket. We observe the strong, clear signal with a negligible contribution from the gasket material. As indicated, the strongest powder-like signal belongs to *hcp*-Fe and MgO. There is an additional contribution from diamond anvils in a form of strong spots. Data was collected at the wavelength of 0.4834 Å.